 \documentclass[final,3p,12pt,times]{elsarticle}

\usepackage{lmodern}
\usepackage{graphicx}
\usepackage{pgf}  
\usepackage{subfig}
\usepackage{amsmath}
\usepackage{longtable}    
\usepackage{morefloats}  
\usepackage{graphicx}
\usepackage{xcolor}
\usepackage{url}
\usepackage{caption}
\usepackage{multicol}
\usepackage{multirow}
\usepackage{array}
\usepackage{lscape}
\usepackage{boxedminipage}
\usepackage{multicol}
\usepackage{multirow}
\usepackage{booktabs}
\usepackage[pagewise]{lineno}
\usepackage[titletoc,title]{appendix}
\usepackage{amssymb}
\usepackage[Symbol]{upgreek}

\usepackage[mathscr]{euscript}  
\DeclareMathAlphabet{\mathscrbf}{OMS}{mdugm}{b}{n}  

\usepackage[nodots,nocompress]{numcompress}

\biboptions{numbers, square}

\newcommand{\eref}[1]{Eq.~\ref{#1}}
\newcommand{\fref}[1]{Fig.~\ref{#1}}
\newcommand{\tref}[1]{Table~\ref{#1}}
\newcommand{\sref}[1]{Section~\ref{#1}}

\journal{}

\begin{document}

\begin{frontmatter}



\title{Symmetry-preserving WENO limiters}

\author{Xiaodong Liu}
\ead{xliu@lanl.gov}
\author[]{Nathaniel R. Morgan\corref{cor1}}
\ead{nmorgan@lanl.gov}
\author{Donald E. Burton}
\ead{burton@lanl.gov}
\cortext[cor1]{Corresponding author}

\address{X-Computational Physics Division; Los Alamos National Laboratory; P.O. Box 1663, Los Alamos, NM, USA}

\begin{abstract}
Weighted essentially non-oscillatory (WENO) reconstruction schemes are presented that preserve cylindrical symmetry for radial flows on an equal-angle polar mesh. These new WENO schemes are used with a Lagrangian discontinuous Galerkin (DG) hydrodynamic method. The solution polynomials are reconstructed using the WENO schemes where the DG solution is the central stencil. A suite of challenging test problems are calculated to demonstrate the accuracy and robustness of the new WENO schemes.
\end{abstract}

\begin{keyword}
Lagrangian shock hydrodynamics \sep Discontinuous Galerkin \sep Cell-centered hydrodynamics \sep Symmetry-preserving WENO
\end{keyword}

\end{frontmatter}


\section{Introduction}  
Lagrangian hydrodynamic methods, \textit{e.g.}, staggered-grid hydrodynamics (SGH) \cite{VNR,WilkinsSGH,BurtonSGH,Caramana,MorganMARS} and cell-centered hydrodynamics (CCH)\cite{Godunov1, Godunov2, Despres, Maire1, Maire2, BurtonCCH,Boscheri,MorganContact}, solve the governing equations for gas (or solid) dynamics on a mesh that moves and deforms with the flow. DG methods \cite{ShutvbDG41990, ShuDG5jcp1998, luodgtaylor2008, LiurDGcaf2017} have been developed for Lagrangian hydrodynamics \cite{JiaDGlagjcp2011,VilarDGlagcaf2012, VilarDGlagjcp2014, LiuDGlagcaf2017,LiebermanStrength1D2017,LiuDGlagRZ2018,MorganVeloFilter2017,LiuDGlagSMS2019}. 
For strong shock problems, both Barth-Jesperson limiter \cite{BJlimiter1989, kuzminbarth2010} and the WENO reconstruction method \cite{AbgralWENO1994,ShuWENO1994,luodghweno2007,QiuHWENO2004} have been explored with Lagrangian DG methods \cite{VilarDGlagcaf2012, VilarDGlagjcp2014, LiuDGlagRZ2018,MorganVeloFilter2017,LiuDGlagSMS2019,JiaDGlagjcp2011}. 
However, the research on symmetry preserving WENO reconstruction schemes is quite limited \cite{ChengCCH2010}, and as such, it is the focus of this paper.

In \cite{LiuDGlagcaf2017}, the modal DG method generates a system of equations to evolve the coefficients for a Taylor series polynomial forward in time.  
The specific volume, velocity, and specific total energy fields are approximated with Taylor series polynomials about the mass center of a reference cell.
The Lagrangian DG hydrodynamic method conserves mass, momentum, and total energy. 
 An explicit TVD Runge-Kutta (RK) method is employed for time marching.  
 In order to preserve cylindrical symmetry for radial flows on an equal-angle polar mesh, new WENO schemes are presented that build the WENO reconstruction by either projecting to a local orthonormal basis \cite{Maire3} or using a local characteristic decomposition \cite{Shuchar1998}.  These WENO schemes are used with a Lagrangian DG method in this work where the DG solution is used as a central stencil.  These WENO schemes could also be used with finite volume CCH methods where the central stencil is constructed by least squares fitting neighboring cell average values.
 
\section{Discretization}
\label{TaylorDG}
The differential Lagrangian equations for the specific volume ($\nu$), velocity (${\textbf {\textit{u}}}$), and specific total energy ($\tau$) evolution are given by,

\begin{equation}
\label{governing}
\begin{array}{lll}
\rho \frac{{d \nu}} {dt}  = \nabla \cdot  {\textbf {\textit{u}}}, & \rho \frac{d{\textbf {\textit{u}}} }{dt}  = \nabla \cdot {{\upsigma}}, &\rho \frac{d \tau}{dt}  = \nabla  \cdot ( {\upsigma}  {\textbf {\textit{u}}} ),
\end{array}
\end{equation}
\noindent where ${\upsigma}$ is the stress tensor.  
The pressure, specific internal energy and specific kinetic energy are denoted as $p$, $e$, and $k$ respectively. 
For gas dynamics, ${\upsigma} = -p \bf I$.  The time derivatives are total derivatives that move with the flow.  
The rate of change of the position is, $\frac{d {\textbf {\textit{x}}} }{dt}  ={{\textbf {\textit{u}}}}$.
Please refer to \cite{LiuDGlagcaf2017} for more details about nomenclatures.
\begin{figure} 
\centering
\includegraphics[trim=0 0 0 0, width=3in,height=!,clip=true]{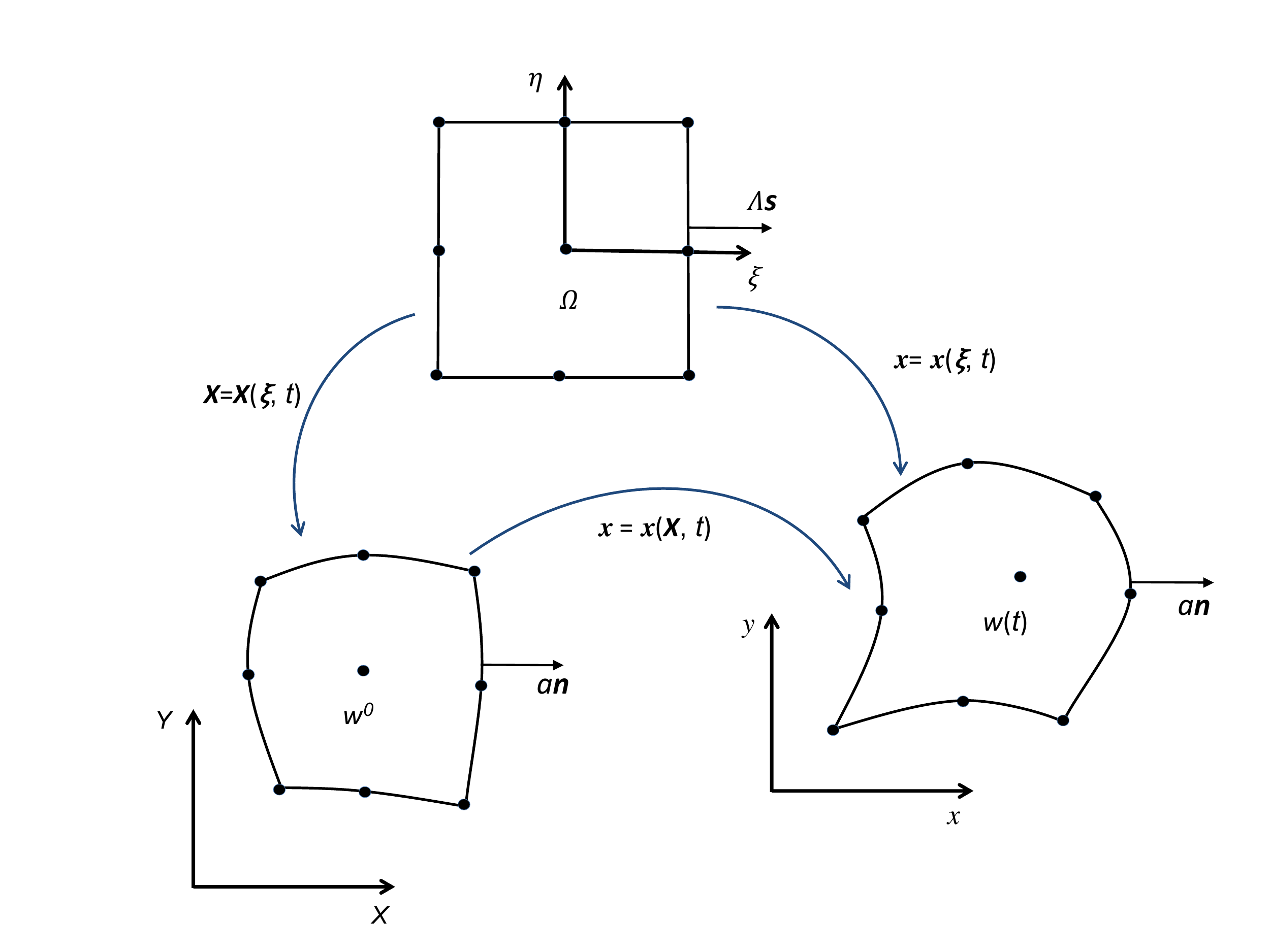}
\caption{\label{NomenclatureFig}  The map from the initial configuration ($X-Y$) and the map from a reference cell ($\xi-\eta$) are graphically illustrated.
The cells will deform with the flow and the volume at a later time will be $w(t)$.  }
\end{figure}

Unknown fields $q$ (\textit{e.g.,} $v$, $\textbf {\textit{u}}$ and $\tau$) can be represented with Taylor expansions on the reference cell $\mathit{\Omega}$ about the center of mass.  
\begin{equation}
\label{TaylorExpansion0}
\begin{array}{lll}
q &= & \sum_{n=1}^{N(P)}q_n \psi_n
\end{array}
\end{equation}
where $q_1 = \bar q_{cm}$, $q_2 =\frac{ \partial q}{\partial \xi}\bigg{|}_{cm}$, $q_3 =\frac{ \partial q}{\partial \eta}\bigg{|}_{cm}$, and 
$\psi_1 = 1$, $ \psi_2 = \xi - \xi_{cm}$, $ \psi_3 = \eta - \eta_{cm}$.
$N(P)$ means the number of terms for the solution polynomial with degree $P$ ( $N(P)$ is equal to 3 for DG(P1)).
The subscript $cm$ denotes the center of mass, given by $(\xi_{cm}, \eta_{cm}) = \frac {1}{m}{\int \limits_{w(t)}\rho (\xi, \eta) dw}$.
Here, $m$ denotes the mass of the cell $w(t)$, namely $m={\int \limits_{w(t)} \rho dw}$.
The basis functions ($\psi_n$) are constant in time \cite{LiuDGlagSMS2019}.
The evolution equation (\textit{e.g.}, the specific volume equation) is multiplied by the Taylor basis functions and then integrated over the current cell configuration.

\begin{equation*}
\label{IntegralSpecificVol}
\begin{array}{lll}
\int \limits_{w(t)}  {\psi}_m \left( \rho \frac{d v }{dt} - \nabla \cdot  {\textbf {\textit{u}}} \right) dw &=& 0 , \qquad  m = 1,...,N(P).
\end{array}
\end{equation*}
Through a set of math operations, the resulting evolution equations for the unknown basis coefficients are,

\begin{equation}
\label{DGEquation-SpecV}
\begin{array}{lll}
\sum_n {\text M}_{mn} \frac{d{v}_n}{dt} = \oint \limits_{\partial w(t)} {\psi}_m ({\textbf {\textit{n}}}\cdot{\textbf {\textit{u}}}^*) da -
\int \limits_{\mathit{\Omega}} {\textbf {\textit{u}}} \cdot
  \nabla 
 {\psi}_m dw, \qquad m = 1,...,N(P).
 \end{array}
\end{equation}

\noindent Here,  ${\text M}_{mn}= \int \limits_{w(t)} \rho {\psi}_m {\psi}_n dw $. The 1st term on the right hand side (rhs) in Eq. (\ref{DGEquation-SpecV}) requires solving a Riemann problem \cite{BurtonCCH} on the surface of the deformed cell $w(t)$.  The Riemann velocity and stress are denoted with a superscript $*$.  
The multidirectional approximate Riemann solver for Lagrangian  CCH has been explored extensively \cite{Despres,Maire1, BurtonCCH}.  
The volume integral (the 2nd term on the rhs) is evaluated by Gauss quadrature formulas. 
An explicit TVD RK method \cite{ShuDG5jcp1998} is used to evolve the semi-discrete system of equations.  

\section{WENO limiting}
\label{Limiters}

The Lagrangian DG hydrodynamic method evolves polynomial coefficients forward in time and these polynomial coefficients must be limited near shocks to ensure monotone solutions.   In this paper, these solution polynomials are reconstructed using two WENO schemes where the DG solution is the central stencil. 
It is very important to preserve cylindrical symmetry with WENO on an equal-angle polar mesh for 1D radial flows. 
To preserve symmetry, two strategies are explored for creating a WENO reconstruction, which are (1) a projection to a local orthonormal basis and (2) a local characteristic decomposition.  These strategies are 
denoted in this paper as strategy 1 and 2.

\subsection{Projection to a local orthonormal basis}
\label{locorth}
  Step 1. We project the DG solution matrix ${\text U}$ in the reference space ($\xi-\eta$) to the physical space ($x-y$) for ${\text U}_p$.
  Here {\text U} is the DG solution matrix, defined by,
     
  \begin{equation}
   \label{solution-matrix}
   {\text U}=
    \begin{bmatrix}
     \bar \nu &  \frac{\partial \nu}{\partial \xi}  & \frac{\partial \nu}{\partial \eta}\\
     \bar  u &  \frac{\partial u}{\partial \xi}  & \frac{\partial u}{\partial \eta}\\
     \bar  v &  \frac{\partial v}{\partial \xi}  & \frac{\partial v}{\partial \eta}\\
     \bar \tau &  \frac{\partial \tau}{\partial \xi}  & \frac{\partial \tau}{\partial \eta}\\
   \end{bmatrix}.
 \end{equation}  
 In the physical space, $ q_p = \sum_{n=1}^{N(p)} q_{pn}\phi_n$
 with  $q_{p1} = \bar q_{p}$, $q_2 =\frac{ \partial q_p}{\partial x}\bigg{|}_{cm}$, $q_3 =\frac{ \partial q_p}{\partial y}\bigg{|}_{cm}$ and $\phi_1 = 1$, $\phi_2 = x -  x_{cm}$ and $\phi_3 = y - y_{cm}$.
Here, $(x_{cm}, y_{cm})$ is the physical mass center for the cell, namely $(x_{cm}, y_{cm}) = \frac {1}{m}{\int \limits_{w(t)}\rho (x, y) dw}$.
The corresponding polynomial  $q_p$ in the physical space can be calculated using  $L_2$ projection, 
          \begin{equation}
             \label{L2proj}
             \begin{array}{lll}
             \Big[\sum_{n=1}^{N(P)} \int \limits_{w(t)}  \rho {\phi}_m \psi_n dw \Big] q_n &=&  \Big[\sum_{n=1}^{N(P)} \int \limits_{w(t)} \rho {\phi}_m \phi_n dw \Big] q_{pn} , \qquad  m = 1,...,N(P).
             \end{array}
         \end{equation}
  Likewise, the polynomial matrix in the physical space, ${\text U}_p$, can be obtained.
  
  Step 2. We construct the polynomials $q_p^r$ from the selected stencils.
\begin{figure} 
\centering
\includegraphics[trim=0 0 0 0, width=2in,height=!,clip=true]{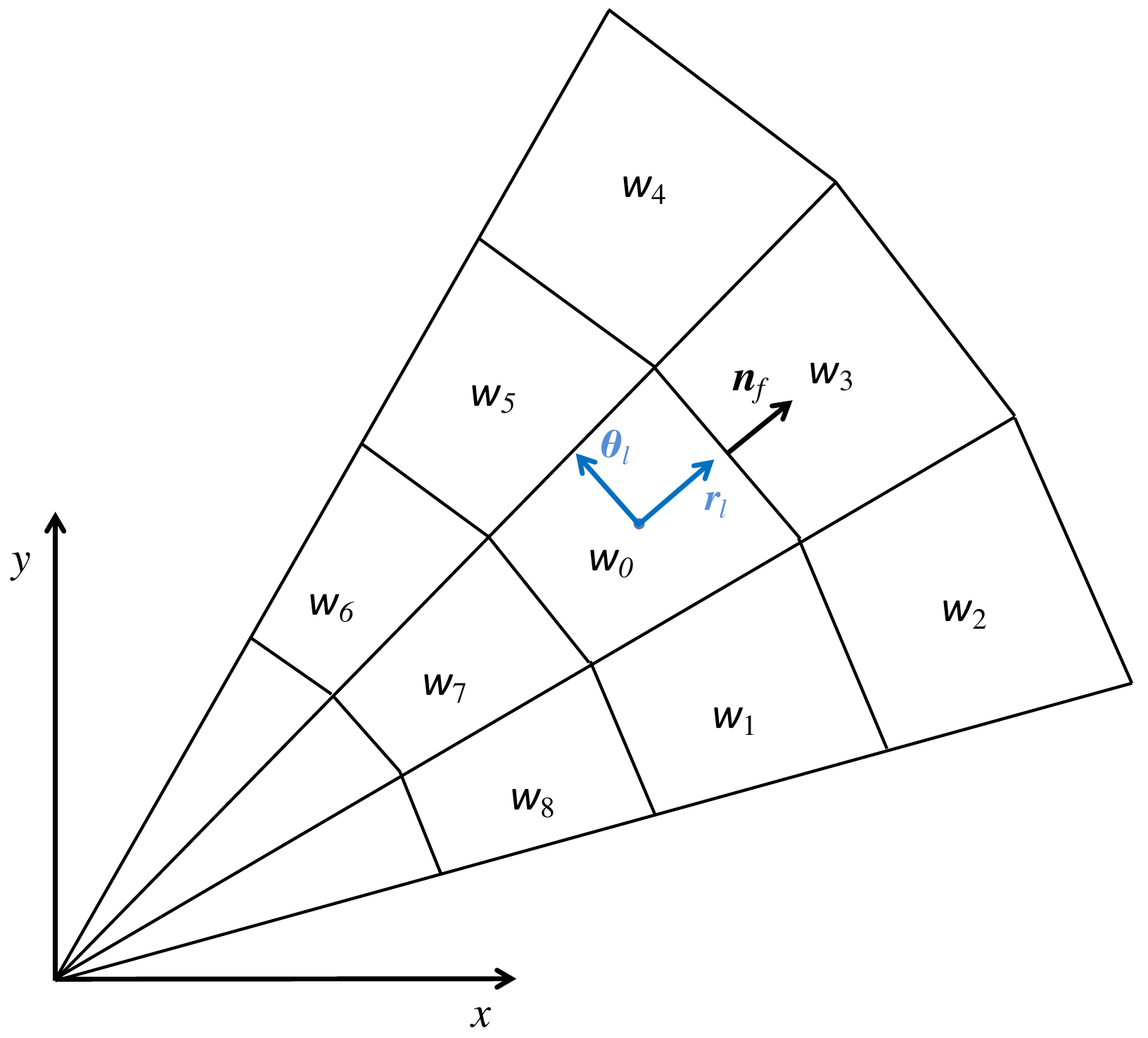}
\caption{\label{Stencil}  The stencils associated with the cell $w_0$.  
${\textbf {\textit{r}}}_l$ and $\boldsymbol{\theta}_l$ denote the local orthonormal basis determined by the 
local cell average velocity. ${\textbf {\textit{n}}}_{f}$ represents the unit normal vector for the the face $f$ surrounding the cell $w_0$. }
\end{figure}
   For the cell $w_0$, shown in \fref{Stencil}, the following 5 stencils, 
    $\{S_0: w_0\}$,  $\{S_1: w_0, w_7, w_8, w_1\}$,   $\{S_2: w_0, w_1, w_2, w_3\}$,   $\{S_3: w_0, w_3, w_4, w_5\}$,   and  $\{S_4: w_0, w_5, w_6, w_7\}$,
    have been selected.
    Here, the polynomial of the central stencil $S_0$ is known, while the polynomials from other 4 biased stencils need to be reconstructed. 
    For every biased stencil, let's assume the reconstructed polynomial is ${q_p^r} ={\bar q}_{p}(w_0) + \frac{\partial q_p^r}{\partial x}(x-x_{cm}(w_0)) + \frac{\partial q_p^r}{\partial y}(y-y_{cm}(w_0))$.
    Taking stencil $S_1$ as an example, this polynomial ${q_p^r}$ satisfies the following,
      \begin{equation}
                \begin{array}{lll}
                 \int \rho q_p^{r} d w_7 = m(w_7) \bar q_p (w_7), &  \int \rho q_p^{r} d w_8= m(w_8) \bar q_p (w_8) & {\text {and}}  \int \rho q_p^{r} d w_1 = m(w_1) \bar q_p (w_1).
                   \end{array}
         \end{equation}      
  Then least squares can be used to calculate  $q_{p2}^r$ and $q_{p3}^r$. Likewise we can get reconstructed polynomial matrix ${\text U}_{p}^r$ for the biased stencils. 

 Step 3. We compute the smoothness indicator.
    The smoothness indicator depends on the variable gradient, that is frame dependent and thus leads to rotational symmetry loss.  
    
    Step 3.1 Project the polynomial matrix ${\text U}_{p}^r$ to a local orthonormal basis for ${\text U}_{pl}$. 
   For the cell $w_0$, we define a local orthonormal basis using the local cell average velocity $(u, v)$, namely
${\textbf {\textit{r}}}_l = \frac{1}{\sqrt{u^2+v^2}} [u \quad v]^T$ and $\boldsymbol{\theta}_l = \frac{1}{\sqrt{u^2+v^2}} [-v \quad u]^T$, shown in \fref{Stencil}. 
Therefore, a transformation matrix is introduced,         
  \begin{equation}
   \label{trnfm-mat}
   {\text B}= 
   \begin{bmatrix}
     \frac{\partial r}{\partial x}&  \frac{\partial r}{\partial y}\\
     \frac{\partial \theta}{\partial x}&  \frac{\partial \theta}{\partial y}
   \end{bmatrix}  
   =
   \frac{1}{\sqrt{u^2+v^2}}
    \begin{bmatrix}
     u &  v\\
    -v &  u
   \end{bmatrix}.
 \end{equation}
We transform all the stencil polynomials ${\text U}_{p}^r$ in the physical space to the local  basis,
 
\begin{equation}
\label{local-basis}
 \left\{
 \begin{array}{ccc}
 \bar{\nu}_{pl} &=&  \bar{\nu}_p^{r}\\
\bar{\tau}_{pl} &=&\bar{\tau}_p^{r}
\end{array},\\
\begin{bmatrix}
   \bar{u}_{pl}\\
   \bar{v}_{pl}
\end{bmatrix} = {\text B}
\begin{bmatrix}
   \bar{u}_p^{r}\\
   \bar{v}_p^{r}
\end{bmatrix},\\
\right.
 \left\{
   \begin{array}{ccc}
   \begin{bmatrix}
       \frac{\partial \nu_{pl}}{\partial r}  &   \frac{\partial \nu_{pl}}{\partial \theta}
   \end{bmatrix} 
   &=& 
   \begin{bmatrix}
     \frac{\partial \nu_p^{r}}{\partial x}  &   \frac{\partial \nu_p^{r}}{\partial y}
   \end{bmatrix} 
  {\text B}^{-1}\\
\begin{bmatrix}
  \frac{\partial \tau_{pl}}{\partial r}  &   \frac{\partial \tau_{pl}}{\partial \theta}
\end{bmatrix} 
&=& 
\begin{bmatrix}
  \frac{\partial \tau_p^{r}}{\partial x}  &   \frac{\partial \tau_p^{r}}{\partial y}
\end{bmatrix} 
{\text B}^{-1}
\end{array}
\right.\\,
\begin{bmatrix}
  \frac{\partial u_{pl}}{\partial r} &  \frac{\partial u_{pl}}{\partial \theta}\\
  \frac{\partial v_{pl}}{\partial r} &  \frac{\partial v_{pl}}{\partial \theta}
\end{bmatrix} = {\text B}
\begin{bmatrix}
  \frac{\partial u_p^{r}}{\partial x} &  \frac{\partial u_p^{r}}{\partial y}\\
  \frac{\partial v_p^{r}}{\partial x} &  \frac{\partial v_p^{r}}{\partial y}
\end{bmatrix}
{\text B}^{-1}.
\end{equation}
where the subscript $l$ denotes variables in the local basis. 

    Step 3.2 Calculate the smoothness indicator in context of a local basis. 
     Then the smoothness indicator $o_{i} $ for a stencil is calculated by,
       \begin{equation}
        \label{Smoothid}
        \begin{array}{lll}
          o_{i} &=& {\text {max}}(o_{i}(\nu_{pl}), o_i(u_{pl}), o_i(v_{pl}), o_i(\tau_{pl}) ), \qquad  i = 0,...,4.
         \end{array}
         \end{equation}
with $o_{i}(q_{pl}) =  \Big(\frac{\partial q_{pl}}{\partial r} \Big)^2+ \Big(\frac{\partial q_{pl}}{\partial \theta}\Big)^2$.
Since the terms $\frac{\partial \nu_{pl}}{\partial r}$ and $\frac{\partial \nu_{pl}}{\partial \theta}$ only depend on radius $r$, the cylindrical symmetry is preserved. 
 
   Step 3.3 We compute the nonlinear weights $\omega_i$ based on the smoothness indicator $o_i$,
       \begin{equation}
        \label{Weight-nl}
        \begin{array}{ll}
          \omega_{i} = \frac{\bar \omega_{i} }{\sum \bar \omega_{i} }, & \bar \omega_{i} =\frac{\omega l_i}{(\varepsilon+o_i)^2}
         \end{array}
         \end{equation}
 where $\omega l_i$ is a linear weight and $\varepsilon=10^{-6}$ to avoid division by zero. In this work, $\omega l_0=0.5$ and  
   $\omega l_i$ for biased stencils is just arithmetic average of $(1-\omega l_0)$. 
   
   Step 3.4. We get the WENO reconstruction polynomial matrix in the local basis using ${\text U}_{pl}^{w} = \sum_{i=0}^{4} \omega_{i} {\text U}_{pli}$.  
   
     Step 3.5. We project the  WENO reconstruction polynomial matrix ${\text U}_{pl}^{w}$  in the local basis back to the physical space for ${\text U}_{p}^{w}$ using the inverse process of \eref{local-basis}.
        
  Step 4. We Project the WENO-based polynomial matrix  $\text U_p^w$ in the physical space back to the reference space for $\text U$ using the inverse process of $L_2$ projection defined in \eref{L2proj}.

  \subsection{ Local characteristic decomposition}
  \label{localchar}
     
     Local characteristic decomposition is also applied. 
     Step 1, 2 and 4 are same as that in \sref{locorth}, while Step 3 is done by a local characteristic decomposition detailed as follows.
     
     Step 3.1 We project the polynomial matrix ${\text U}_{p}^r$ in the physical space to the characteristic field for ${\text U}_{c}$.
    The Jacobian matrix of the integral governing equations (\eref{governing}) is, 
     
 \begin{equation}
 \label{Jacobian-mat}
\begin{bmatrix}
   0 &  -n_x & -n_y & 0\\
  -\rho p n_x &  -(\gamma-1)\rho u n_x & -(\gamma-1)\rho v n_x & (\gamma-1)\rho n_x\\
  -\rho p n_y &  -(\gamma-1)\rho u n_y & -(\gamma-1)\rho v n_y & (\gamma-1)\rho n_y\\
  -\rho p v_{n} &  -(\gamma-1)\rho u v_{n} + pn_x & -(\gamma-1)\rho v v_{n} + pn_y  & (\gamma-1)\rho v_{n}\\
\end{bmatrix},
\end{equation}
where  $v_{n} = un_x + vn_y$. This matrix admits 4 eigenvalues, $\lambda_1 = -\rho c$, $\lambda_2 =\lambda_3 =  0$ and $\lambda_4 = \rho c$. 
The left and right eigenvectors of such a matrix are,

 \begin{equation}
 \label{Left-mat}
 {\text L} = 
\begin{bmatrix}
  -\frac{1}{2\gamma} &  \frac{n_x}{2\rho c}-\frac{(\gamma-1)u}{2\gamma p}& \frac{n_y}{2\rho c}-\frac{(\gamma-1)v}{2\gamma p}&\frac{\gamma-1}{2\gamma p}\\
   \frac{1}{\gamma} &  -\frac{n_y v_{m}}{p}-\frac{u}{p \gamma}& \frac{n_xv_{m}}{p}-\frac{v}{p \gamma} & \frac{1}{p \gamma}\\
  0 &  \frac{n_y}{p} & -\frac{n_x}{p} & 0\\
   \frac{1}{2\gamma} &  \frac{n_x}{2\rho c}+\frac{(\gamma-1)u}{2\gamma p}& \frac{n_y}{2\rho c}+\frac{(\gamma-1)v}{2\gamma p}&-\frac{\gamma-1}{2\gamma p}\\
   \end{bmatrix} \quad \text{and} \quad
 {\text R} = 
\begin{bmatrix}
  -1&  \gamma-1 & (\gamma-1)v_{m} & 1\\
 \rho c n_x &  0 & p n_y & \rho c n_x\\
 \rho c n_y &  0 &-p n_x & \rho c n_y\\
 p+\rho c v_{n} &  p & 0 &  -p+\rho c v_{n} \\
\end{bmatrix}.  
\end{equation}
Here, $v_{m} = v n_x - un_y$ and $c$ is the sound speed. We project the polynomial matrix  ${\text U}_{p}^r$ to the characteristic field, 
   \begin{equation}
     \label{Left-mat}
    {\text U}_{cf}={\text L_f} {\text U}_p^r, \qquad  f = 1,...,4.
    \end{equation}
  where the subscript $f$ represent the face number. For ${\text L}_f$ (or ${\text R}_f$), $n_x$ and $n_y$ are defined using each normal vector of the 4 faces (\fref{Stencil}) and the variables are 
  obtained using the arithmetic average of averaged variables of the 2 cells sharing the same face.
  
  Step 3.2 We calculate the smoothness indicator based on ${\text U}_{cf}$.   
   Taking the first characteristic variable $\nu_{cf}$ as an example, 
   
  \begin{equation}
 \label{nuchar} 
 \begin{array}{llll}
   \bar \nu_{cf}  &=& -\frac{1}{2\gamma} {\bar \nu}+  [\frac{n_x}{2\rho c}-\frac{(\gamma-1)u}{2\gamma p}] {\bar u}+ [\frac{n_y}{2\rho c}-\frac{(\gamma-1)v}{2\gamma p}]{\bar v}+ \frac{\gamma-1}{2\gamma p}{\bar \tau}\\
   \frac{ \partial {\nu_{cf}}}{\partial x} &=& -\frac{1}{2\gamma} \frac{ \partial {\nu}}{\partial x} +  [\frac{n_x}{2\rho c}-\frac{(\gamma-1)u}{2\gamma p}] \frac{ \partial u}{\partial x}+ [\frac{n_y}{2\rho c}-\frac{(\gamma-1)v}{2\gamma p}]\frac{ \partial v}  {\partial x}+ \frac{\gamma-1}{2\gamma p}\frac{ \partial {\tau}}{\partial x}\\
   \frac{ \partial {\nu_{cf}}}{\partial y} &=& -\frac{1}{2\gamma} \frac{ \partial {\nu}}{\partial y} +  [\frac{n_x}{2\rho c}-\frac{(\gamma-1)u}{2\gamma p}] \frac{ \partial u}{\partial y}+ [\frac{n_y}{2\rho c}-\frac{(\gamma-1)v}{2\gamma p}]\frac{ \partial v}{\partial y}+ \frac{\gamma-1}{2\gamma p}\frac{ \partial {\tau}}{\partial y}\\
\end{array}.  
\end{equation}   
For sake of brevity, the superscript $r$ and subscript $p$ are omitted. Combining with \eref{trnfm-mat}, 
  \begin{equation}
  \label{nuchar2} 
 \begin{array}{lll}
    \Big(\frac{ \partial {\nu_{cf}}}{\partial x}\Big)^2 + \Big(\frac{ \partial {\nu_{cf}}}{\partial y}\Big)^2 &=& \Big( d_1\frac{ \partial {\nu}}{\partial x} +  d_2\frac{ \partial {u_{pl}}}{\partial x}   +  d_3\frac{ \partial {\tau}}{\partial x}\Big)^2 +
    \Big( d_1\frac{ \partial {\nu}}{\partial y} +  d_2\frac{ \partial {u_{pl}}}{\partial y}   +  d_3\frac{ \partial {\tau}}{\partial y}\Big)^2\\
    &=& \Big( d_1\frac{ \partial {\nu}}{\partial r} +  d_2\frac{ \partial {u_{pl}}}{\partial r}   +  d_3\frac{ \partial {\tau}}{\partial r}\Big)^2 +
    \Big( d_1\frac{ \partial {\nu}}{\partial \theta} +  d_2\frac{ \partial {u_{pl}}}{\partial \theta}   +  d_3\frac{ \partial {\tau}}{\partial \theta}\Big)^2
  \end{array}
\end{equation}
with $d_1 = -\frac{1}{2\gamma} $, $d_2 = \frac{1}{2\rho c} -\frac{(\gamma-1)\sqrt{2k}}{2\gamma p}$, $d_3=\frac{\gamma-1}{2\gamma p}$ and $u_{pl}=un_x+vn_y$.
 It can be observed that $\Big(\frac{ \partial {\nu_{cf}}}{\partial x}\Big)^2 + \Big(\frac{ \partial {\nu_{cf}}}{\partial y}\Big)^2$  only depend on the radius, thus it preserves the cylindrical symmetry. 
   Step 3.3 and 3.4 are similar to the above section. The above 4 steps are executed four times (4 faces) for creating the WENO reconstruction.
  
   Step 3.5 We transform  the 4 reconstructed polynomial ${\text U}_{cf}^w$ in the characteristic field back to the physical space for ${\text U}_{pf}^w$ using the inverse process
   of \eref{Left-mat}, namely, 
   \begin{equation}
     \label{Right-mat}
     \begin{array}{lll}
    {\text U}_{pf}^{w}&=&{\text R_f} {\text U}_{cf}^w, \qquad  f = 1,...,4.
    \end{array} 
    \end{equation}

   Step 3.6 The final sole polynomial ${\text U}_{p}^w$ for the physical space is just the arithmetic average of the 4 polynomials ${\text U}_{pf}^w$.

\section{Test problems}
\label{TestCases}
In this section, a suite of challenging test problems
are calculated to demonstrate the accuracy and robustness of the WENO-based Lagrangian DG hydrodynamic method.  The test problems are the polar Sod \cite{VilarDGlagcaf2012}, Sedov \cite{sedov} and Noh \cite{Noh}, which all use a gamma-law equation of state (EOS). Some important parameters are listed in \tref{tab:parameter}. 
The initial conditions $(\rho^{0}, u^{0}, v^{0},p^{0})$ for polar Sod (left), Sedov (middle) and Noh (right) are
given by,

\begin{table*}[t]
\begin{center}
\caption
{Important parameters for different cases}
\label{tab:parameter}
\mbox{
\begin{tabular}{l c c c c c}
\hline
 & Gas constant ($\gamma$) & Computational domain & Mesh resolution & Final time\\                    
\hline
 Sod shock tube    & 1.4                & $[0.01, 1]\times[0, 2{\pi}]$   &  $99 \times 48$      &  0.2\\
 Sedov blast             &  1.4               & $[-1.2, 1.2]\times[-1.2, 1.2]$&  $60 \times 60$    & 1.0\\
Noh implosion         &  5/3               & $[-1.0, 1.0]\times[-1.0, 1.0]$&  $100 \times 100$ & 0.6\\
\hline
\end{tabular}}
\end{center}
\end{table*}
\begin{equation*}
\label{Initial}
 \left\{
 \begin{array}{cc}
(1, 0, 0, 1) ,& r < 0.5\\
(0.125, 0, 0, 0.1), & r > 0.5
\end{array},\\
\right.
 \left\{
 \begin{array}{cc}
(1.0, 0, 0, (\gamma-1)\frac{\rho_oE_o}{w_o}),&$Origin$ \\
(1.0, 0, 0, 10^{-6}),&$Others$
\end{array},\\
\right.
 \left\{
 \begin{array}{c}
(1.0, \frac{-x}{\sqrt{x^2+y^2}}, \frac{-y}{\sqrt{x^2+y^2}},10^{-6})
\end{array},\\
\right.
\end{equation*}

\noindent where $r=\sqrt{x^2+y^2}$ for the polar Sod problem.  With the Sedov problem, 'Origin' means the cells containing the origin and $w_o$ denotes the cell volume 
and $E_o=0.244816$.

\begin{figure*}[h!]
\centering
\subfloat[The final mesh]{
\includegraphics[width=1.3in]{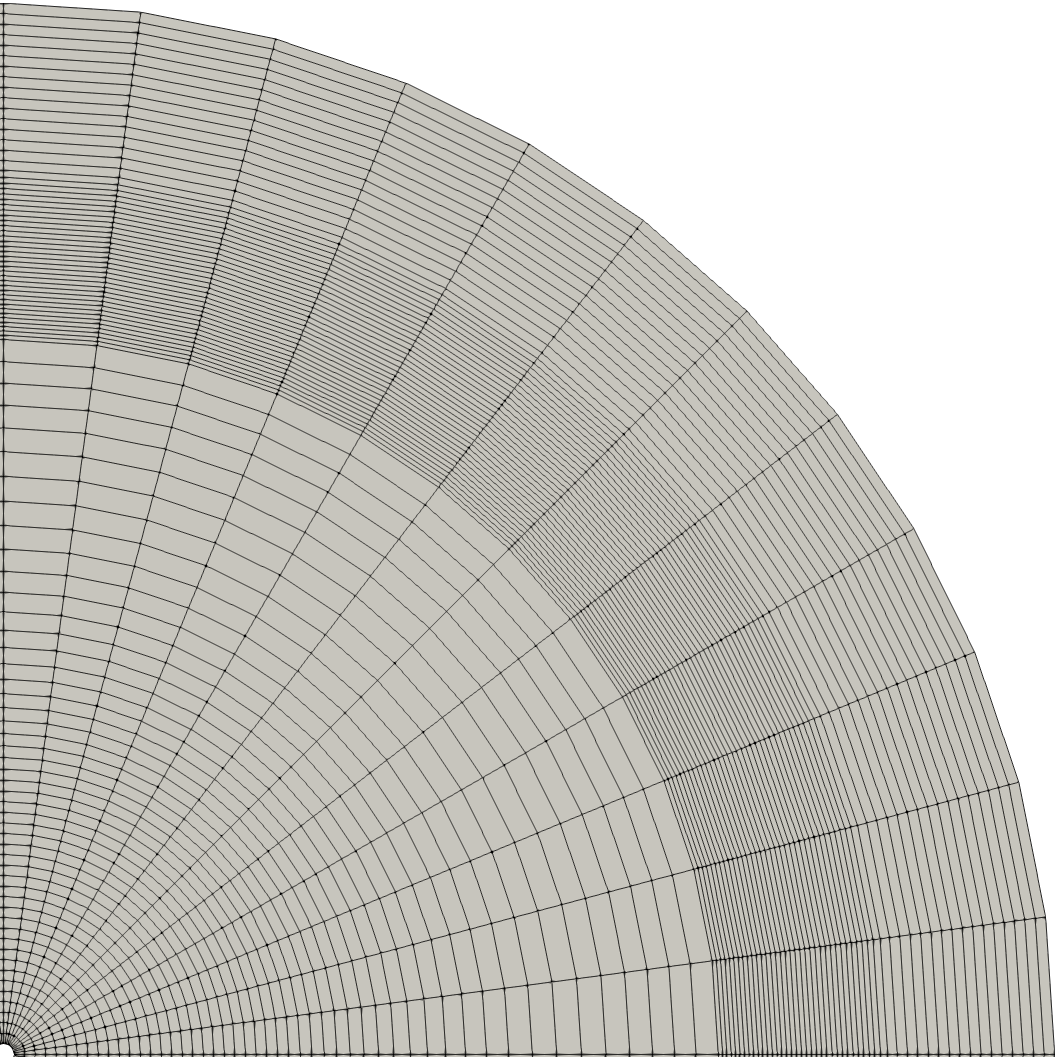}\label{fig:sodmeshsym2}}
\subfloat[Scatter plots of density]{
\includegraphics[width=2.in]{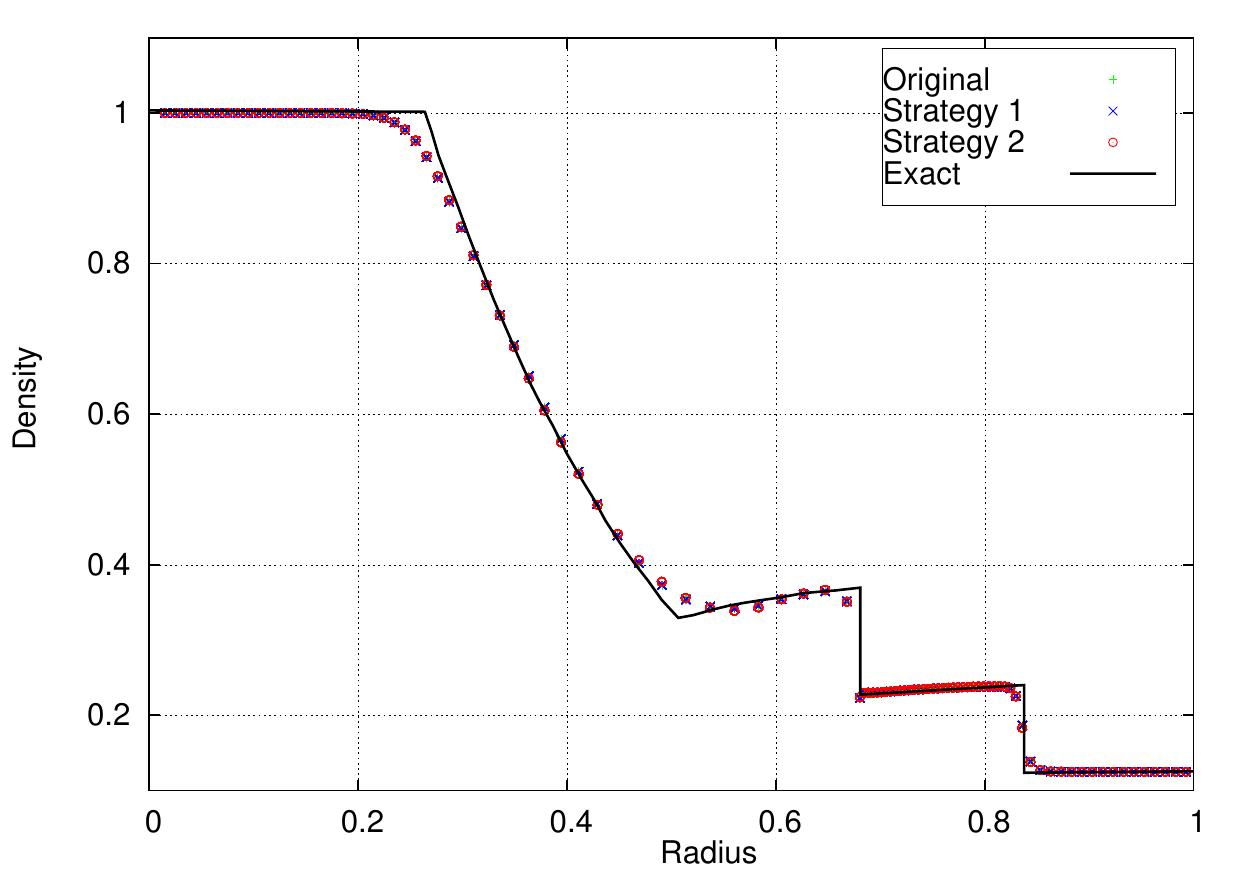}\label{fig:sodrho}}
\subfloat[Scatter plots of symmetry errors]{
\includegraphics[width=2.in]{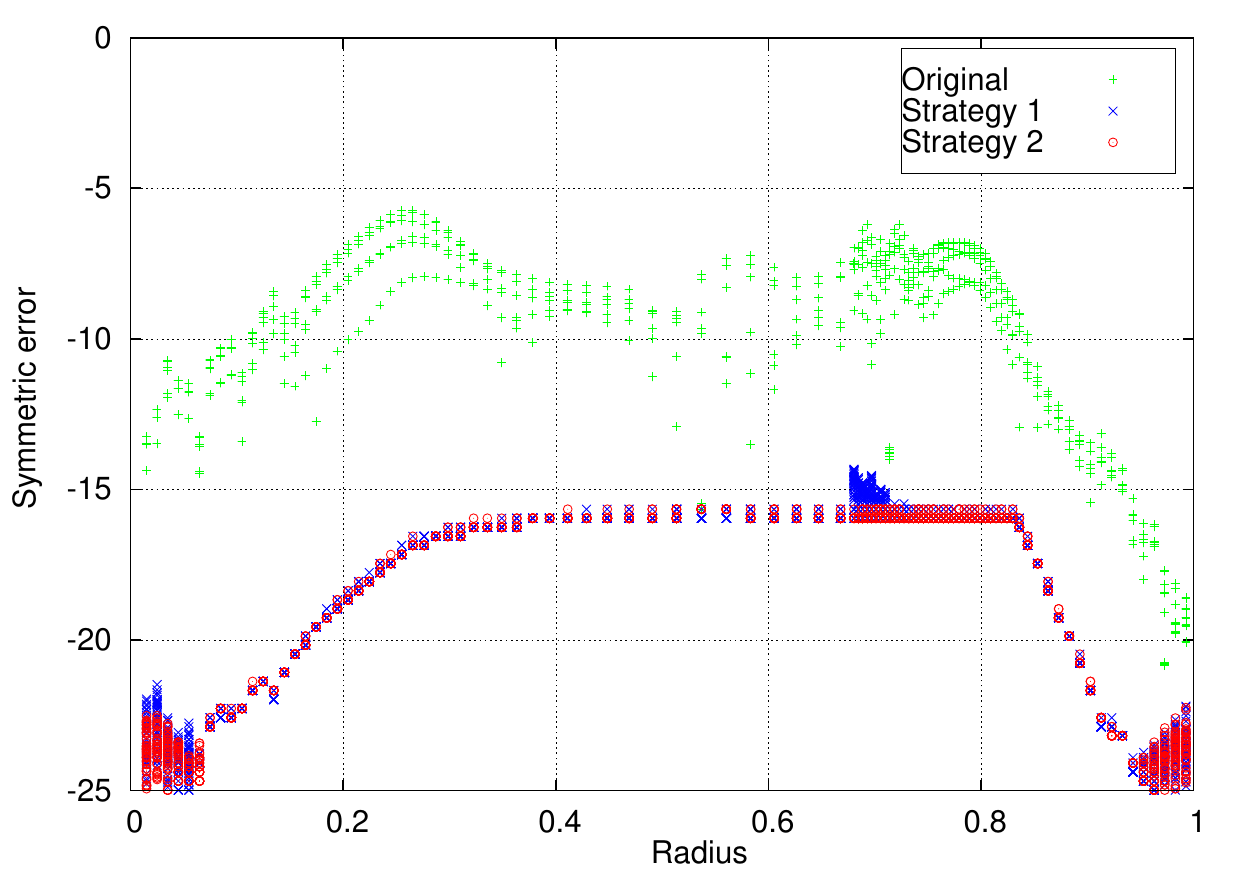}\label{fig:sodreve}}\\
\subfloat[The final mesh]{
\includegraphics[width=1.3in]{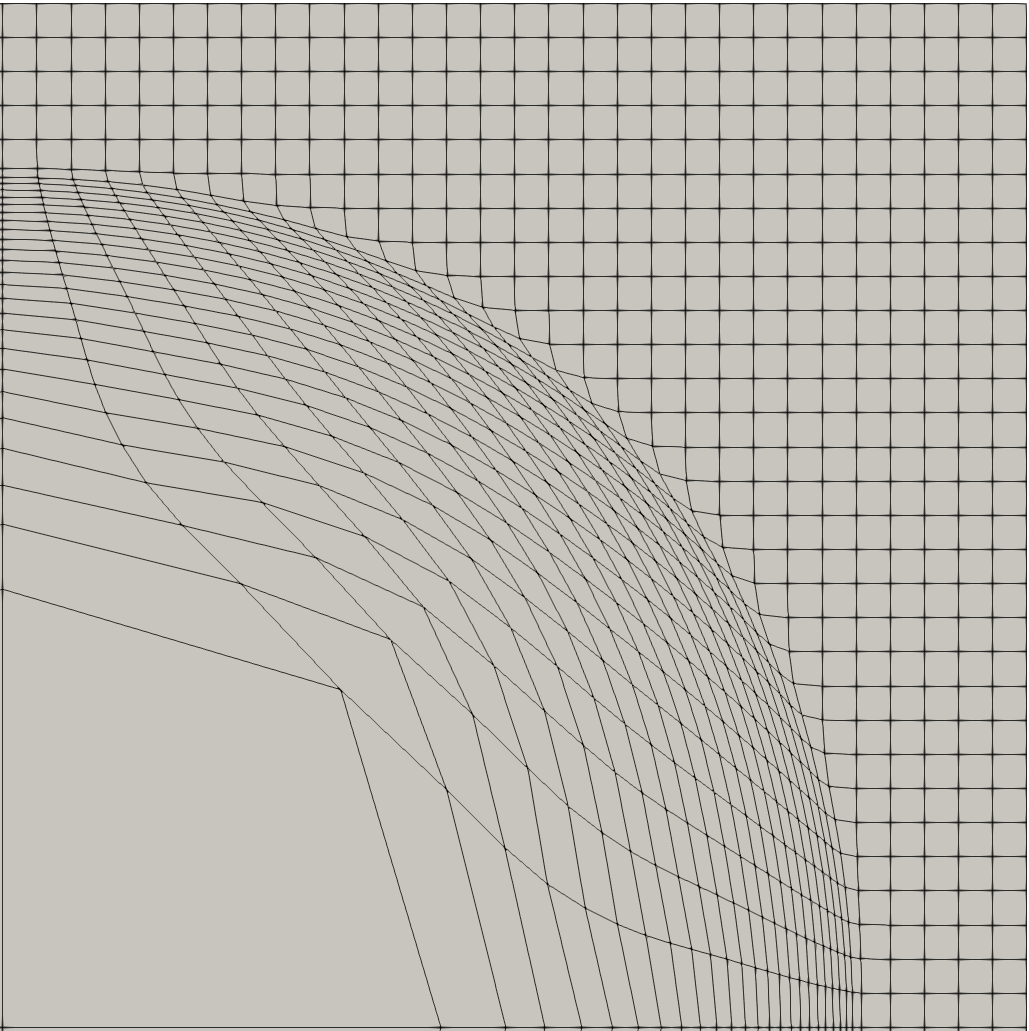}
\label{fig:sedovmeshsym2}}
\subfloat[Scatter plots of density]{
\includegraphics[width=2.in]{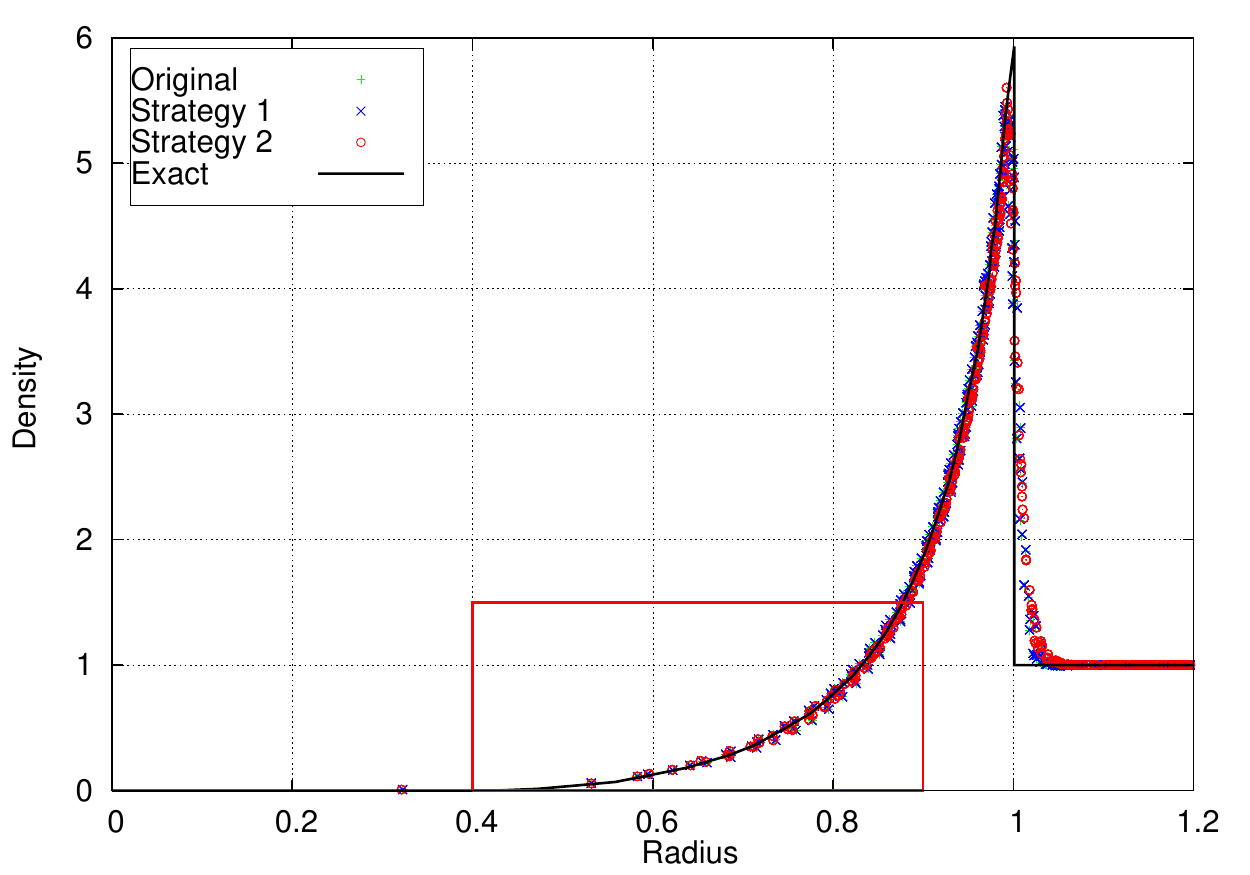}
\label{fig:sedovrho}}
\subfloat[Close-up for the density plots]{
\includegraphics[width=2.in]{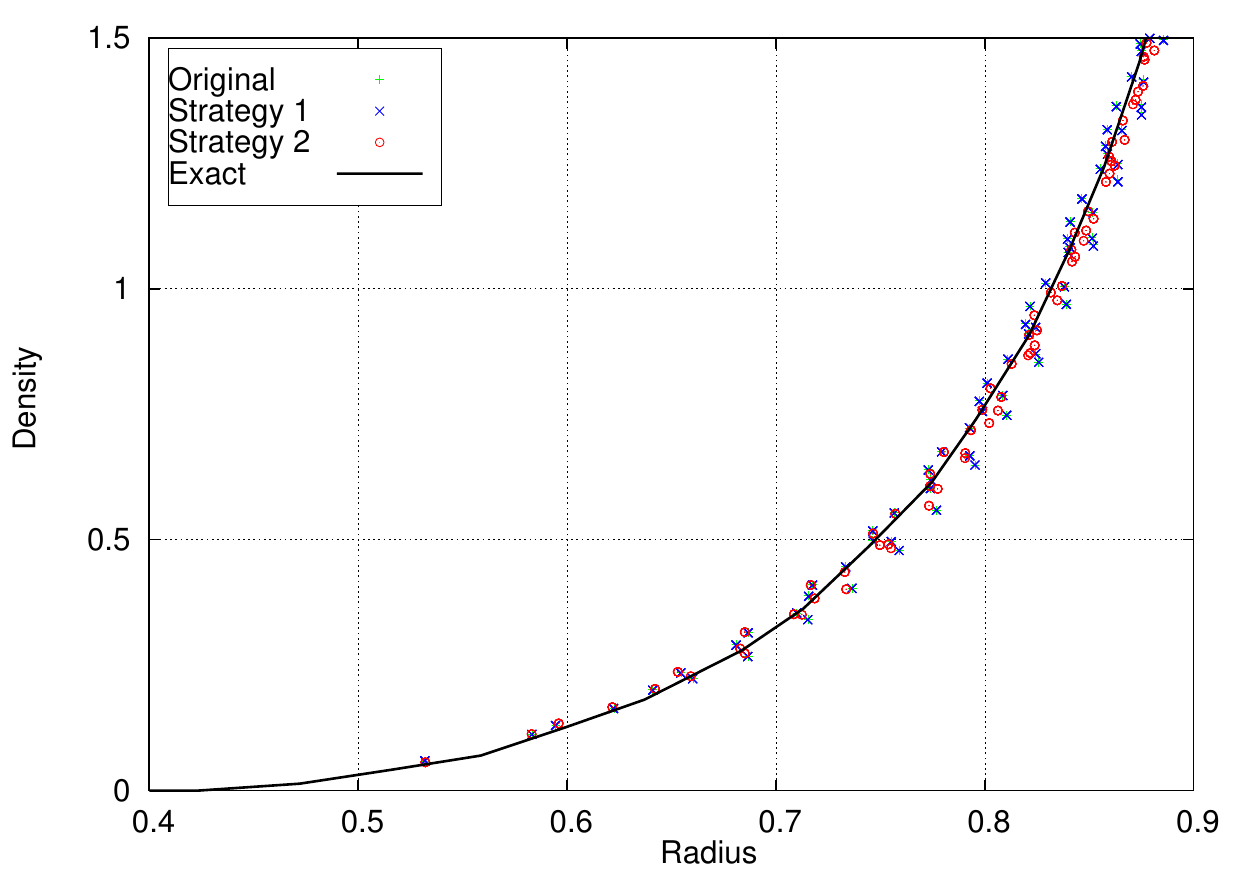}
\label{fig:sedovrho-cu}}\\
\subfloat[The final mesh]{
\includegraphics[width=1.3in]{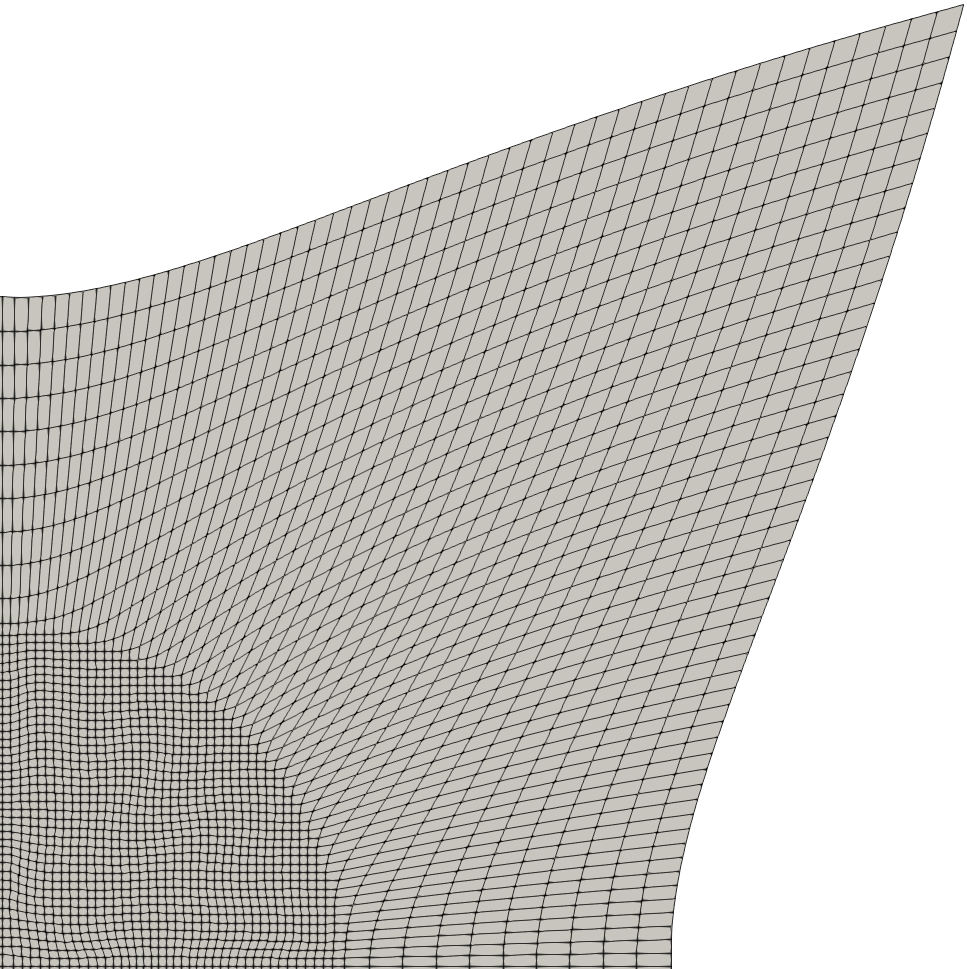}
\label{fig:nohmeshsym2}}
\subfloat[Scatter plots of density]{
\includegraphics[width=2.in]{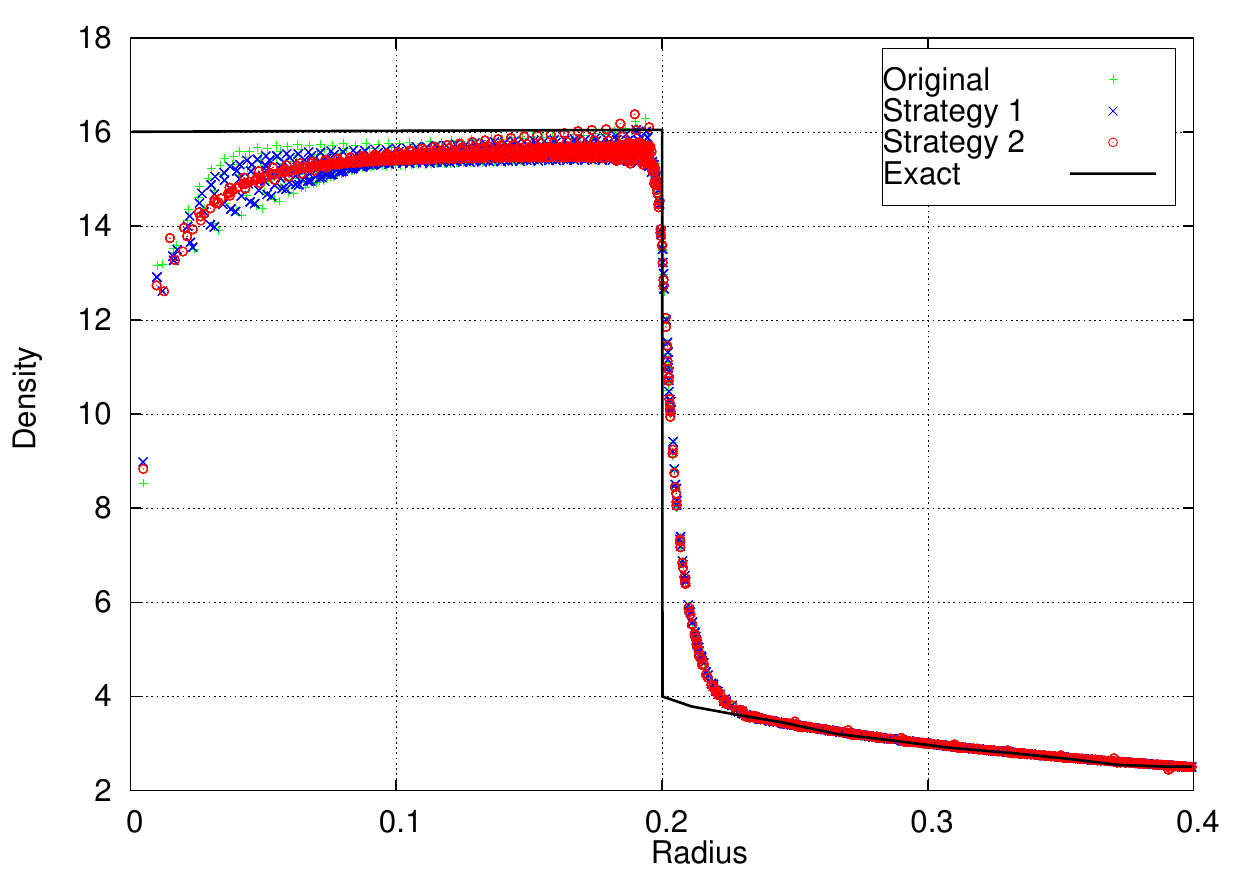}
\label{fig:nohrho}}
\caption{ 
The numerical results are shown for different cases. 
The 1st, 2nd, and 3rd row show the numerical results for the Sod shock tube, Sedov blast wave and Noh implosion test problems respectively.}
\label{fig:results}
\end{figure*}

The numerical results are shown in \fref{fig:results}. 
For all test cases, the meshes in the first quadrant are presented. 
The final meshes (\fref{fig:sodmeshsym2}, \ref{fig:sedovmeshsym2} and \ref{fig:nohmeshsym2}) move in  a stable manner. 
The density scatter plots (\fref{fig:sodrho}, \ref{fig:sedovrho} and \ref{fig:nohrho}) agree well with the exact solution.  
The polar Sod problem is used to do the quantitative analysis of symmetry preserving. 
Velocity deviations from the radial direction are used as a symmetry-preserving metric.
From the scatter plots for symmetry errors shown in \fref{fig:sodreve}, our two strategies preserve symmetry very well as demonstrated by the fact that the largest errors are on the order of $10^{-15}$ (machine precision), while the original WENO method cannot preserve symmetry. From \fref{fig:sedovrho-cu}, for Sedov, strategy 2 gives the smallest density scatter as expected. In addition, the difference between the original WENO method and strategy 1 is very small 
because the original method happens to preserve symmetry on an initially uniform square mesh for this kind of symmetric explosion problem; the explanation is as follows.
Taking the first quadrant as an example, the test problem and the mesh are both symmetric about the $45^0$ 
line so the smoothness indicators for the two symmetric cells about the $45^0$ line are equal. Likewise, a similar phenomenon can be observed for the Noh problem. 

\section{Conclusions}
\label{Conclusion}

We presented new symmetry-preserving WENO limiters and used them with a Lagrangian DG hydrodynamic method to simulate shock driven flows in 2D Cartesian coordinates.  
The WENO reconstructions were calculated using two approaches (1) a projection to a local orthonormal basis or (2) using a local characteristic decomposition to preserve cylindrical symmetry on an equal-angle polar mesh for radial flows.  The DG solution is used as the central stencil for the WENO reconstructions in this work; however, these WENO schemes could also be used with finite volume hydrodynamic methods where the central stencil is constructed by least squares fitting neighboring cell average values.  The symmetry preservation of the new WENO schemes with the Lagrangian DG hydrodynamic method was demonstrated by calculating the polar Sod problem.  The canonical WENO method breaks symmetry, while the new WENO schemes have errors on the order of machine precision.  The accuracy and robustness of the proposed WENO schemes was then demonstrated by calculating the Sedov and Noh test problems.  These new WENO schemes are a promising approach for calculating limited reconstructions for use with finite volume and DG hydrodynamic methods.

\section{Acknowledgments}

We gratefully acknowledge the support of the NNSA through the Laboratory Directed Research and Development (LDRD) program at Los Alamos National Laboratory.  The Los Alamos unlimited release number is LA-UR-19-22578. 

%
\bibliography{Lag-ref} 
\bibliographystyle{unsrt}

\end{document}